\documentclass{egpubl}
\usepackage{eurovis2025}

\SpecialIssuePaper         

\CGFccby

\usepackage[T1]{fontenc}
\usepackage{dfadobe}  

\usepackage{cite}  
\BibtexOrBiblatex
\electronicVersion
\PrintedOrElectronic
\ifpdf \usepackage[pdftex]{graphicx} \pdfcompresslevel=9
\else \usepackage[dvips]{graphicx} \fi

\usepackage{egweblnk}
\usepackage{xspace}
\usepackage{mdframed}
\usepackage{wrapfig}
\usepackage{xcolor}
\usepackage{tikz}
\usepackage{tcolorbox}
\usepackage{amsmath}
\usepackage{ulem}
\usepackage{multirow}
\usepackage{booktabs} 
\usepackage{array}
\usepackage{orcidlink}

\newcommand{\revised}[1]{\textcolor{black}{#1}}

\newcommand{\tool}{\textsc{VizTA}\xspace}

\newcommand{\eg}{\textit{e.g.}}

\definecolor{myblue}{HTML}{D5EAFA}
\definecolor{myorange}{HTML}{FFDFC7}
\definecolor{mygreen}{HTML}{CEEEE1}
\definecolor{myyellow}{HTML}{FFE8A7}
\definecolor{mypurple}{HTML}{E7D7FF}
\definecolor{mygray}{HTML}{F5F5F5}
\definecolor{lel}{RGB}{125,0,125}

\newcommand{\smallpar}[1]{({\small #1})}

\title[\tool]%
      {\tool: Enhancing Comprehension of Distributional Visualization with Visual-Lexical Fused Conversational Interface}

\author[Wang et al.]{
  \parbox{\textwidth}{\centering 
    Liangwei Wang$^{1}$\orcidlink{0000-0003-3481-3993}, 
    Zhan Wang$^{1}$\orcidlink{0000-0003-3318-6473}, 
    Shishi Xiao$^{3}$\orcidlink{0009-0008-0262-5289}, 
    Le Liu$^{4}$\orcidlink{0000-0002-9758-6620}, 
    Fugee Tsung$^{1,2}$\orcidlink{0000-0002-0575-8254}, 
    and Wei Zeng$^{1,2}$\thanks{Wei Zeng is the corresponding author.}\orcidlink{0000-0002-5600-8824}
  }
  \\
  \parbox{\textwidth}{\centering 
    $^{1}$The Hong Kong University of Science and Technology (Guangzhou), Guangzhou, China\\
    $^{2}$The Hong Kong University of Science and Technology, Hong Kong, China\\
    $^{3}$Brown University, Providence, USA\\
    $^{4}$Northwestern Polytechnical University, Xi'an, China
  }
}

\begin{document}


\maketitle


\begin{abstract}
Comprehending visualizations requires readers to interpret visual encoding and the underlying meanings actively. 
This poses challenges for visualization novices, particularly when interpreting distributional visualizations that depict statistical uncertainty.
Advancements in LLM-based conversational interfaces show promise in promoting visualization comprehension.
However, they fail to provide contextual explanations at fine-grained granularity, and chart readers are still required to mentally bridge visual information and textual explanations during conversations.
Our formative study highlights the expectations for both lexical and visual feedback, as well as the importance of explicitly linking these two modalities throughout the conversation.
The findings motivate the design of \tool, a visualization teaching assistant that leverages the fusion of visual and lexical feedback to help readers better comprehend visualization. 
\tool features a semantic-aware conversational agent capable of explaining contextual information within visualizations and employs a visual-lexical fusion design to facilitate chart-centered conversation.
A between-subject study with 24 participants demonstrates the effectiveness of \tool in supporting the understanding and reasoning tasks of distributional visualization across multiple scenarios.


\end{abstract}

\section{INTRODUCTION}

\begin{figure}[t]
  \centering
  \includegraphics[width=0.99\linewidth]{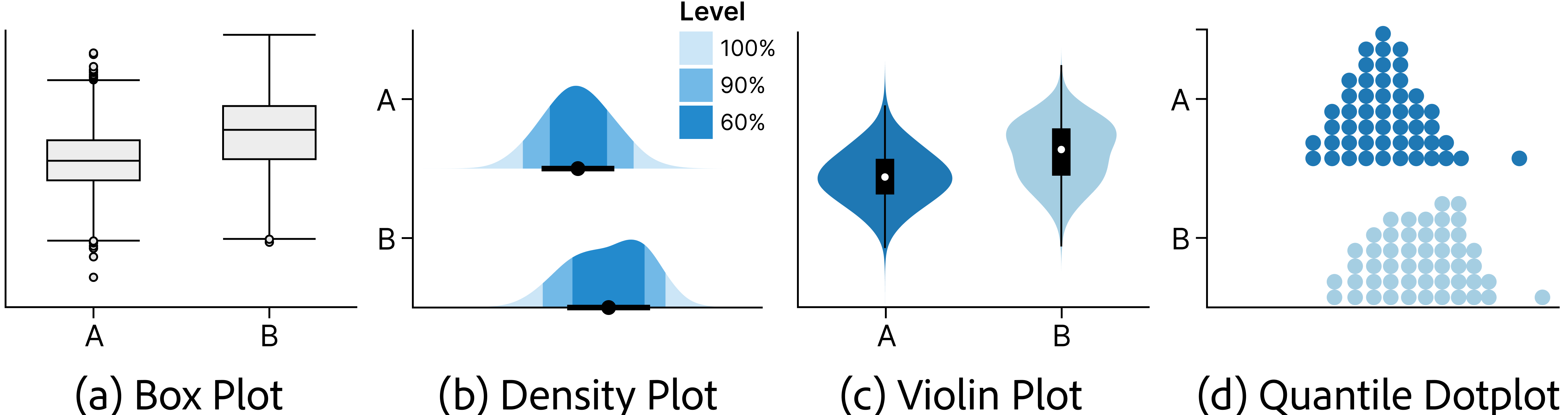}
  \caption{Four types of distributional uncertainty visualizations: box plot, density plot, violin plot, and quantile dotplot.}
   \label{fig:scope}
   \vspace{-5mm}
\end{figure}

Visualization has long been served as an effective bridge for communicating data-driven facts to the public~\cite{vaidya2020knowing}.
However, comprehending complex visualizations, especially those involving statistical uncertainty, remains a challenge not only for the general public but also for policymakers and researchers~\cite{boukhelifa2017data}.
Even highly trained scientists could misinterpret error bars, a widely used uncertainty representation~\cite{belia2005researchers}.
Several factors contribute to this difficulty.
First, the design of uncertainty visualization is diverse, including vast types of representation such as summary displays~\cite{cumming2007error}, continuous marks~\cite{makowski2019bayestestr}, and discretized design through frequency farming~\cite{kay2016ish}.
Fig.~\ref{fig:scope} illustrates four examples of these visualizations. 
Second, the semantic richness of visual symbols in these visualizations poses a unique challenge. 
Even visually similar or identical symbols can carry distinct meanings depending on the data context. 
For instance, an error bar might represent a range, standard deviation (SD), standard error (SE), or confidence interval (CI), requiring readers to distinguish between descriptive and inferential aspects~\cite{cumming2007error}.
Furthermore, uncertainty visualizations are prone to misinterpretation, often unconsciously, due to their tricky visual representations and varying levels of readers' literacy~\cite{ruginski2016non}.
Assisting non-expert readers in understanding uncertainty visualizations remains challenging.

Conversational interaction powered by LLMs offers a promising avenue for visualization education.
Choe et al.~\cite{choe2024enhancing} illustrated how conversational interactions between novice chart readers and LLMs can provide on-demand support for understanding visualizations.
Their approach employed annotations to resolve ambiguities in visual communication, enabling users and the system to share annotated charts during the conversation.
However, existing efforts overlook the unique challenges of conversation interaction centered on visualization.
First, the potential questions from readers span a broad range. 
These questions often focus on either visual elements or data facts, requiring explanations that encompass not only the overall chart but also its fine-grained visual elements.
This demands a conversational agent capable of understanding \textit{multi-granular} visual elements in the chart and providing \textit{context-aware} responses.
Second, interchanges concerning visualizations often rely on referential expressions (known as \textit{deixis}) such as \textit{“this point”}~\cite{han2024deixis}.
The complexity of deixis usage in interchanges concerning visualizations often surpasses what can be effectively conveyed through text accompanied with annotated charts.
When explaining visualizations with multiple references, readers need explicit guidance to navigate between different visual elements in a structured sequence to reduce the cognitive burden of matching textual descriptions with visual information~\cite{Ottley2019TheCC}. 
Nevertheless, existing conversational interfaces for visualization lack effective mechanisms for textual-visual cross-referencing in both users' queries and system responses, hindering effective visual communication.

\revised{This paper presents \tool, a prototype visual-lexical fused conversational interface to aid in visualization reading.}
\tool is informed by a formative study that highlights the challenges general readers face when interpreting distributional visualizations, and the forms of assistance they expect.
\tool features novel \textit{visual-lexical fusion interaction} in the interactive conversation, powered by a \textit{semantic-aware conversational agent} enhanced by multi-source structured data and a citation tutorial.
Finally, we conduct a between-subject user study to assess the effectiveness of \tool.
Quantitative and qualitative results suggest that \tool helps readers better comprehend and reason with distributional visualization. In summary, our key contributions are as follows:

\begin{itemize}

\item We conduct a formative study to identify readers' needs in conversational interface for visualization, and summarize the visual granularity and semantic context in distributional visualization. 

\item We present \tool
, a visualization teaching assistant that offers interactive reading modules featuring visual-lexical fusion interaction design and a semantic-aware conversational agent. 

\item We demonstrate the effectiveness of \tool in aiding readers' comprehension and reasoning performance through a between-subject user study.

\end{itemize}
\section{BACKGROUND AND RELATED WORK}

\subsection{Uncertainty in Distributional Visualization}

Extensive works have explored efficient design for uncertainty visualization.
Among these, graphical annotation and visual encoding are two broad techniques to present uncertainty~\cite{padilla2020uncertainty}.
Graphical annotation uses marks to display distributional properties, including summary plot (\eg, error bars~\cite{cumming2007error, correll2014error}, \revised{box plots~\cite{krzywinski2014visualizing}}), continuous distributional plot (\eg, density plots~\cite{makowski2019bayestestr}), and discretized representations (\eg, quantile dotplot~\cite{kay2016ish}, icon arrays~\cite{garcia2013communicating}).
Visual encoding depicts uncertainty in visual channels (\eg, fuzziness, size, transparency) that are perceptually difficult to quantify.
Another type of work emphasizes analyzing how readers perceive and understand uncertainty visualization.
They point out that readers often have misconceptions when reading uncertainty visualization\cite{ruginski2016non, joslyn2021visualizing}, and human perception (\eg, size effect~\cite{kale2020visual}, visual saliency of annotation~\cite{padilla2017effects}) can also affect the inferring process. 
Despite efforts to make uncertainty visualization more accessible for non-expert users through creation tools~\cite{kay2023ggdist} or enhance statistical report readability~\cite{masson2023statslator}, there remains a gap in assisting them in interpreting these charts effectively. 
Faced with challenges in interpreting uncertainty visualization, there is a natural inclination to explore how to assist the public in comprehending it, echoing the call for visualization education~\cite{bach2023challenges}.

\textbf{Scope of Our Work.} 
A variable with uncertainty cannot be adequately described by a single deterministic value. 
To represent the variability effectively, it is necessary to consider features from its potential distributions,  such as using \{$\mu$, $\sigma$\}  for normally distributed variables. 
The grammar in ggdist~\cite{kay2023ggdist} introduces an effective strategy for visualizing uncertainty by mapping distribution functions to visual aesthetics.
Accordingly, we communicate uncertainty through the lens of distributional visualization.
The uncertainty visualizations chosen for this work include box plots, density plots, violin plots, and quantile dotplots (or hybrids with interval bars), as shown in Fig.\ref{fig:scope}. 
\revised{The reasons are twofold.
First, these visualizations offer versatility and clear visual distinctions for 1D uncertainty and are commonly used in real-world scenarios while presenting a meaningful level of comprehension difficulty.
Second, they utilize graphical annotation rather than visual encoding, making them more conducive to natural language explanations that complement the visualization to reduce cognitive load~\cite{bancilhon2023combining}.}

\subsection{Natural Language Interface for Chart Comprehension}

There is a recognized variation of individual comprehension capability on charts, termed \textit{visualization literacy}~\cite{wileman1993visual}, with established assessment methods~\cite{lee2016vlat,pandey2023mini}.
Natural language assistance~\cite{shen2023towards} and multi-modal interaction~\cite{srinivasan2017orko} have emerged as an effective strategy to bridge the literacy gap, with many visualization comprehension tools building upon this principle.
Inksight~\cite{lin2023inksight} utilizes sketch input to summarize insights that align with user intent.
\revised{
AutoVizuA11y~\cite{duarte2024autovizua11y} and Vizability~\cite{gorniak2024vizability} convert visual information from charts into playable text through keyboard navigation, enhancing chart accessibility.
To support network visualization education, Shu et al.~\cite{shu2024particular} introduced interactive pattern explanation, enabling users to select specific visualization areas for automated pattern mining and visual-textual explanations.
Closely related to our work,} Choe et al.~\cite{choe2024enhancing} demonstrated the value of combining visual and textual access in visualization communication, showcasing the potential of LLM-assisted approaches in visualization education.
However, existing conversational interfaces for charts has key limitations: the lack of contextual information for discussing fine-grained elements, and textual explanations are not explicitly linked to visual features, making the navigation of conversation records challenging.
\revised{Our work advances this direction by introducing the idea of visualization-text interplay and forming visual-lexical fusion design in conversational interface for visualization.}

\subsection{Visualization-Text Interplay}

Visualization-text interplay has earned significant attention in the field of visualization, with studies examining the design patterns~\cite{latif2021deeper} and cognitive impact~\cite{zhi2019linking,Ottley2019TheCC}.
Numerous studies suggest that fine-grained integration of visualization and text, where text serves beyond mere chart descriptions, can significantly enhance communication effectiveness.
For instance, \revised{for data-rich paragraph comprehension,} Charagraph~\cite{masson2023charagraph} \revised{and GistVis~\cite{zou2025gistvis}} generate interactive charts with bidirectional visualization-text highlighting. 
\revised{To support interactive document creation,} VizFlow~\cite{sultanum2021leveraging} enables dynamic layouts in data-driven articles in scrollytelling, and Kori~\cite{latif2021kori} constructs interactive references between text and visualizations created with Vega-Lite. 
\revised{Recently, Cai et al.~\cite{cai2024linking} have also proposed a semi-automatic pipeline using knowledge graphs for authoring text with text-visualization linking. 
For data-driven writing,} EmphasisChecker~\cite{kim2023emphasischecker} coordinates highlights between line charts and captions, and DataTales~\cite{sultanum2023datatales} leverages LLM to generate textual narratives with given charts.
Similar approaches could facilitate visualization-centered communication in remote meetings~\cite{han2024deixis}, \revised{visual analytics in NLG-based interface~\cite{srinivasan2018augmenting}, automatic graphical overlays design~\cite{hao2024finflier}}, and even the creation of narrative data videos~\cite{shen2024data, wang2024wonderflow}.
Despite the effectiveness, visualization-text linking remains mainly unexplored in conversational visualization interfaces.

\section{BRIDGING VISUAL AND LEXICAL UNDERSTANDING}

We conducted a formative study to understand the difficulties general readers face with distributional visualizations and the support they need.
Building on these findings, we analyzed the granularity and semantics of visual elements in distributional visualizations as a foundation for bridging visual and lexical understanding.

\subsection{Formative Study}
\textbf{Procedure and Tasks.} 
We recruited 9 participants (4 males and 5 females) \revised{through convenience sampling from a local university.
Most participants (N=8) self-reported being familiar with common visualizations (bar, pie, line charts, etc.), but were not fully versed in uncertainty visualizations. 
One participant reported minimal prior exposure to reading visualizations.}
We did not impose restrictions on their backgrounds, as we hope to get inspiration from people with varying levels of visualization literacy rather than relying on experts.
After introducing our research topic to participants, we provided four types of uncertainty visualizations as stimuli, consistent with Fig.~\ref{fig:scope}. 
Each chart had a distinct data context and descriptive text outlining its background.
When reading these charts, participants were asked to demonstrate their process of understanding in a think-aloud protocol.
They were encouraged to voice any questions or doubts about the information displayed on the visualization. 
After articulating their comprehension and doubts, participants were asked to envision how an AI system could assist them in understanding visualizations and the feedback they expected.
Throughout the process, we asked follow-up questions if a response was unclear or if we wished to delve deeper.

\noindent
\textbf{Finding: Reading strategies for distributional visualization.}
Chart readers have distinct objectives at each stage when reading visualizations, as outlined by Walny et al.~\cite{walny2017active}: recognizing, tracking, reorganizing, decoding, and analyzing. 
In the context of distributional visualization, we observed three key stages in participants' active reading process. 
They formulated different questions at each stage based on their specific understanding goals.

\begin{itemize}
\item \textit{Recognizing: Seeking General Overview.}
A general overview can act as metadata for the chart that assists readers in quickly engaging with the narrative context~\cite{burns2022invisible}.
It entails both the structure of the chart, including its type and arrangement of key axes, and its contextual semantics shaped by the data.

\item \textit{Decoding: Understanding Visual Elements.} Decoding involves interpreting information from visual channels like color, shape, size, and position.
While this initial decoding identifies the visual elements, a comprehensive understanding requires recognizing the statistical meaning encoded in thes.
For instance, readers might see \textit{``95\% CI''} with endpoints at $x_1$ and $x_2$, while a deeper understanding is needed to interpret it as \textit{``we can be 95\% confident the true population mean falls within the range [$x_1$, $x_2$]''}.

\item \textit{Analyzing: Comparing, Reasoning and Inferring.}
Advanced analysis in chart reading goes beyond interpreting presented data to generate new insights through comparison, reasoning, and inference~\cite{brehmer2013multi}.
This process integrates visual information to draw conclusions, make decisions, and assess uncertainties, \eg, deciding to buy coffee while waiting for a bus~\cite{kay2016ish}.
\end{itemize}

\noindent
\textit{Design Goal:}
There is a broad spectrum of potential questions readers may raise, spanning from visual details to underlying data facts shown on charts.
To support this, we prioritize natural language interaction in our system design, enabling readers to freely pose questions throughout each stage of their chart-reading process.

\noindent
\textbf{Finding: Multi-modal queries and multi-granular deixis usage.}
Natural language expressions often failed to preserve the conversational context for certain questions targeting chart details.
In such cases, participants tended to combine deixis with pointing gestures to convey their intent, as in queries like, \textit{``What do the x and y values of [this point] indicate?''} or involving multiple visual entities like \textit{``How does [this] compare to [that]?''}
We observed that \textit{deixis} usage in \textit{queries} spans varying complexities, including one-to-one, one-to-many, and many-to-one mappings, aligning with findings by Han et al.~\cite{han2024deixis}.
Moreover, when discussing distributional visualizations, participants naturally decomposed chart elements across different levels of abstraction for reference.
We identified two primary granularities of visual elements commonly referenced through deixis.
The first type involves visual components such as points, lines, or areas within the chart. 
The second type is based on the semantic meaning conveyed in data, where deixis refers to one data category or group represented in the visualization.

\noindent
\textit{Design Goal:}
It is essential to incorporate multi-modal interaction mechanisms that support natural language with visual inputs to streamline the conversation. 
The varying complexities of deixis usage highlight the necessity of enabling repeated and flexible visual references. 
We define these two granularity levels of deixis-based visual element referencing as \textit{element-level} and \textit{group-level} to establish a common ground for the visual interaction (\S\ref{ssec:granularity_semantic}).

\begin{table*}[t]
\centering
\footnotesize
\begin{tabular}{>{\centering\arraybackslash}p{2.2cm}|>{\centering\arraybackslash}p{0.8cm}|>{\centering\arraybackslash}p{2.5cm}|>{\centering\arraybackslash}p{2.2cm}|p{7.7cm}}
\toprule
\textbf{Type} & \textbf{\#Focus} & \textbf{Numerical Encoding} & \textbf{Examples} & \textbf{Example Templates for Semantic Context}\\
\midrule
\multirow{2}{*}{Summary Mark} & 1 & $\{v\}$ & Outlier Dot \includegraphics[width=0.24cm]{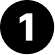} & The point shows an outlier at \{v\}.\\
& 2 & $\{v_1, v_2\}$ & IQR Box \includegraphics[width=0.24cm]{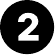} & The box span from $\{v_1\}$ to $\{v_2\}$, indicating an IQR with $\{v_2-v_1\}$.\\
\midrule
\multirow{2}{*}{Continuous Mark} & \multirow{2}{*}{$\geq$3} & $v(x)$, $x \in X$ & Density Area \includegraphics[width=0.24cm]{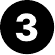}  & The area shows a density distribution, spanning from $\{x_{start}\}$ to \{$x_{end}$\}, and has \{peak(s)/trough(s)\} at \{$x_{peak}/x_{trough}$\}.\\
& & $v(x)$, $x \in [x_1,x_2]$ & Truncated Density Area \includegraphics[width=0.24cm]{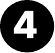}  & \textit{The area shows a density distribution, with an interval from} $\{x_1\}$ to $\{x_2\}$, indicating \{role\}.\\
\midrule
Discretized Mark & 3 & $\{v_1, v_2, v_3\}$ & Dot Bin \includegraphics[width=0.24cm]{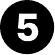}  & The dot bin accounts for approximately $\{v_1\}$ of the total sample, centered at $\{v_2\}$. 
It also indicates a one-sided cumulative probability $P(X \leq \{v_2\}) \approx \{v_3\}$ from start.\\
\midrule
Functional Mark & 0 & $\emptyset$ & Legend \includegraphics[width=0.24cm]{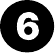}  & The legend indicates \{role\}.\\
\bottomrule
\end{tabular}
\caption{The table displays the four classifications of element-level visual marks, categorized by the numerical encoding pattern. 
\#Focus denotes the number of numerical data points encoded. Examples of the visual elements mentioned are highlighted in Fig.~\ref{fig:granularity}a.}
\label{tab:visual_encoding}
\vspace{-3mm}
\end{table*}

\noindent
\textbf{Finding: Preferences for fine-grained, contextual and human-like response.}
Participants desired explanations that are contextual and deeply connected to the data presented in the chart.
For example, they wanted to understand not only the visual encoding method but also interpret its meaning in the data context.
They also favored explanations that integrate examples derived directly from the visualization to enhance clarity and relevance, as suggested by Lundgard et al.~\cite{lundgard2022accessible}.
In addition, when presented with support, participants wanted guidance that directs their attention to specific elements on the chart related to their queries rather than just providing dry textual descriptions. 
As one participant envisioned, \textit{``like a face-to-face conversation, pointing out what I need to focus on without me searching.''}
Moreover, participants emphasized the importance of seamlessly integrating textual explanations with visual guidance. 
They envisioned the AI system acting as a thoughtful storyteller, delivering clear and engaging textual narratives while dynamically guiding attention to relevant chart elements as needed. 

\noindent
\textit{Design Goal: }
Participants emphasized the importance of reading guidance in two modalities: (1) textual explanation and (2) visual guidance.
To better integrate visual and textual information, we propose a visual-lexical fusion design, which incorporates visual cues directly into textual narratives (\S\ref{ssec:fusion_design}).
To provide fine-grained, contextual explanations, we design a semantic-aware conversational agent tailored for chart interpretation (\S\ref{ssec:agent}).

\begin{figure}[t]
  \centering
  \includegraphics[width=0.99\linewidth]{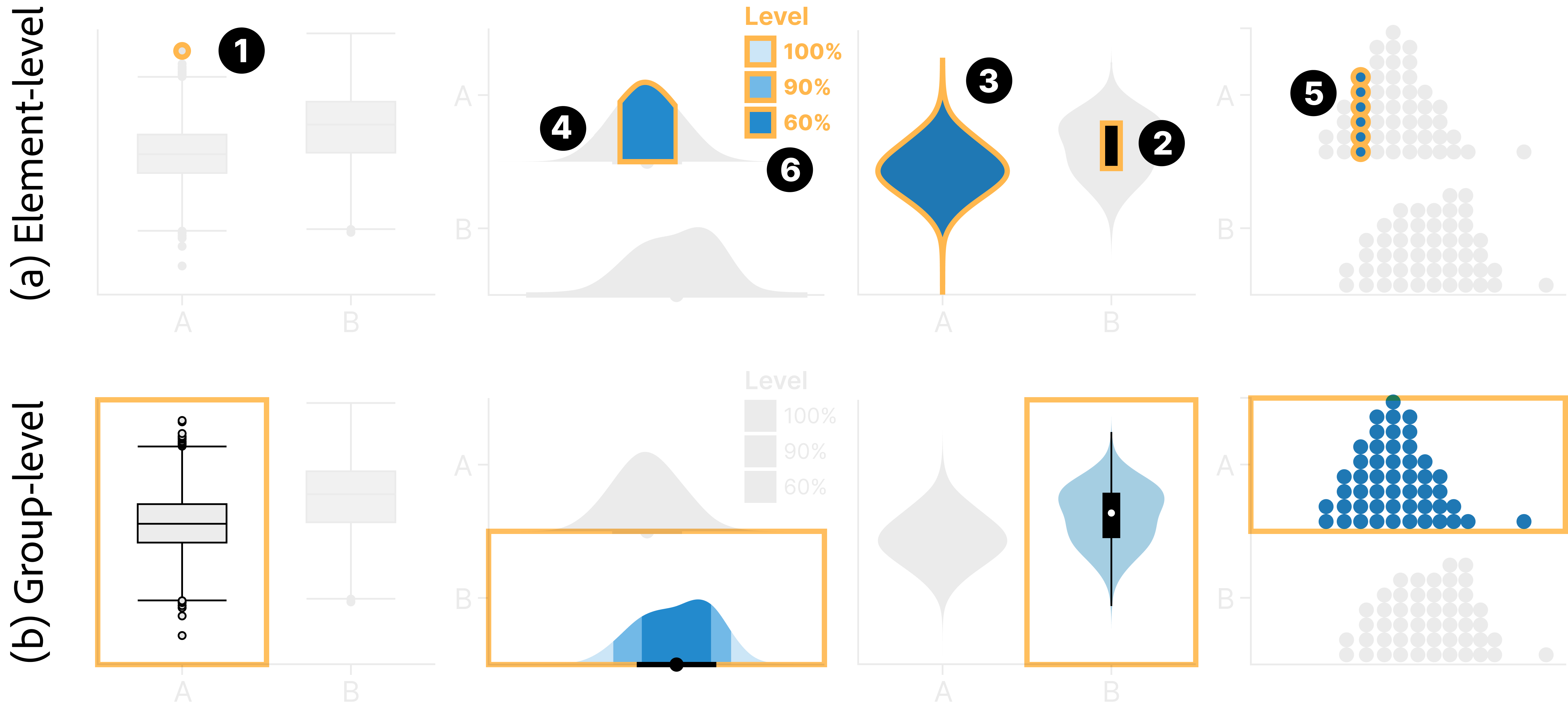}
  \caption{Examples of visual elements at element-level and group-level granularity in distributional visualizations.}
   \label{fig:granularity}
   \vspace{-6mm}
\end{figure}

\subsection{Visual Granularity and Semantic Context}
\label{ssec:granularity_semantic}

Our formative study reveals that users naturally reference visualization elements at different granularities during their reading process.
However, such references are often unsystematic, potentially leading to ambiguous or meaningless visual deixis.
To establish clear mappings between visual representations and their statistical meanings, we define two levels of granularity as \textbf{\textit{element-level}} and \textbf{\textit{group-level}}, and then systematically extract their associated semantic context.
This summarization provides a foundation for generating accurate natural language explanations in subsequent steps.

\textit{Element-level} marks are the fundamental building blocks of visualizations (Fig.~\ref{fig:granularity}a), including both the basic visual elements (i.e., points, lines, and areas) and functional elements that provide interpretative frameworks.
\revised{By analyzing their distinct encoding patterns for numerical data, we categorize these \textit{element-level} marks into four types and determine the semantic context for all mark types in our discussed visualizations, as shown in Table \ref{tab:visual_encoding}.}

\begin{itemize}
\item \textit{Summary marks} encode individual data points or value pairs to abstract distribution characteristics, for example, an outlier dot \includegraphics[width=0.30cm]{src/fig/icon/granularity1.pdf} in box plots or the interquartile range (IQR) \includegraphics[width=0.30cm]{src/fig/icon/granularity2.pdf} in violin plots.

\item \textit{Continuous marks} represent continuous distribution information through smooth visual expressions, requiring multiple data points for construction, as commonly seen in density plots \includegraphics[width=0.30cm]{src/fig/icon/granularity4.pdf}.
\item \textit{Discretized marks} represent distributional with discrete visual elements through frequency framing~\cite{fernandes2018uncertainty}, which bridges the gap between single-value representation and continuous distribution, exemplified by dot bins \includegraphics[width=0.30cm]{src/fig/icon/granularity5.pdf} in quantile dotplots~\cite{kay2016ish}.
\item \textit{Functional marks} serve an interpretative role without directly encoding data, providing reference frameworks such as axes and legends \includegraphics[width=0.30cm]{src/fig/icon/granularity6.pdf} that support the understanding of other marks.
\end{itemize}

\textit{Group-level} marks provide a full view of the distribution by aggregating multiple element-level marks that share the same categorical encoding (Fig.~\ref{fig:granularity}b). 
For instance, one violin group represents an entire distribution through density curves \textit{(continuous mark)} and integrating elements that show quartile statistics \textit{(summary marks)}. 
\section{\tool}
\label{sec:vista}

In this section, we first present an overview of \tool (\S\ref{ssec:system_overview}).
Then we introduce the visual-lexical fusion interaction design (\S\ref{ssec:fusion_design}) and the construction of the backbone \revised{LLM-based} semantic-aware conversational agent (\S\ref{ssec:agent}) in \tool.

\subsection{System Overview}
\label{ssec:system_overview}
\begin{figure}[t]
  \centering
  \includegraphics[width=0.99\linewidth]{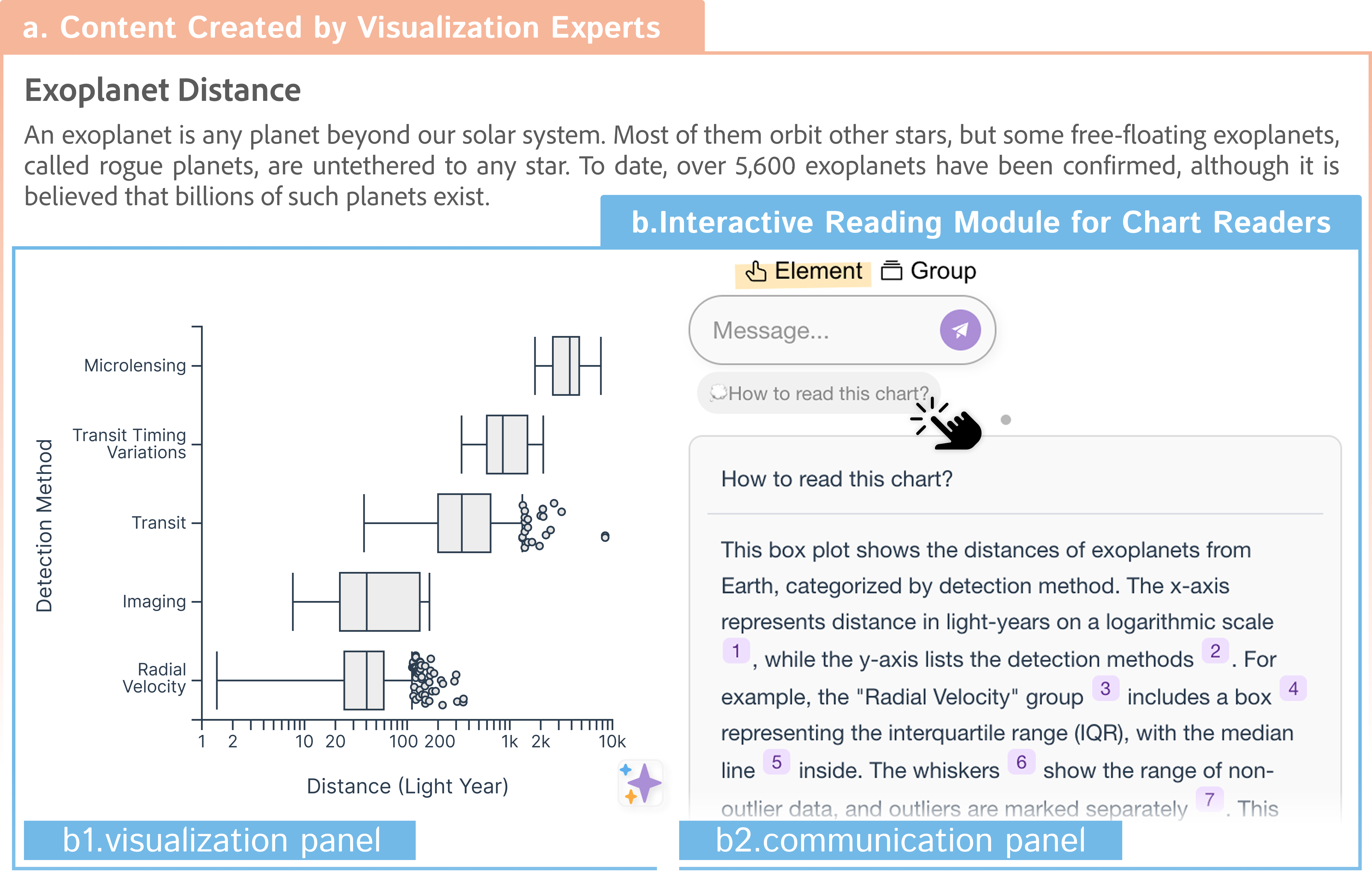}
  \caption{\label{fig:overview}
           Inspired by AutoVizuA11y~\cite{duarte2024autovizua11y}, real-world chart reading involves two primary user groups: visualization experts, responsible for creating and publishing charts, and chart readers, the end users who benefit from the reading assistance.
           \vspace{-3mm}
    }
\end{figure}

\revised{\tool is a web-based tool that supports visualization experts in creating distributional visualizations and enables a visual-lexical fused conversational interaction to help chart readers form better comprehension.}
Designed as a Vue component, \tool allows visualization experts to create charts in web content by uploading data in CSV format and configuring chart parameters (Fig.~\ref{fig:overview}a).
\revised{In the chart generation process, \tool automatically constructs a prompt to initialize the LLM-based conversational agent that is capable of communicating the intricate details of the visualization.}

As shown in Fig.~\ref{fig:overview}b, \tool serves as an interactive reading module for chart readers.
Optimized for on-demand access, \tool displays only the visualization panel (Fig.~\ref{fig:overview}b1) by default. 
When readers click the \includegraphics[width=0.30cm]{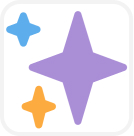} button to activate \tool, the communication panel (Fig.~\ref{fig:overview}b2) appears and anchors at right, with a message box for users to input their queries.
Buttons for adjusting interaction granularity are located above the message box, while system-recommended prompts are displayed below for quick access.
Readers' queries can incorporate visual elements in \textit{element-level} or \textit{group-level} granularity by performing drag-and-drop actions directly on the chart. 
Responses from \tool are concise and presented in small cards to maintain clarity and prevent information overload.
Textual explanations in responses are enhanced with inline citations, represented by \includegraphics[width=0.30cm]{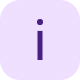} (i=1, 2, 3...), which, when hovered over, highlight the corresponding visual element and display tooltip in the visualization panel.
Each card encapsulates a single-turn dialogue, displaying both the query and response. 
As the conversation progresses, additional cards stack sequentially, and readers can navigate the conversation log using small dots at the top of each card.

\subsection{Visual-Lexical Fusion Design}
\label{ssec:fusion_design}
To ensure efficient and smooth visual communication, we introduce visual-lexical fusion design for visualization conversational interaction.
It features two key mechanisms: incorporating multi-granular chart components into queries through direct manipulations, and providing visual-textual context in responses with inline citations.
\revised{
We define the complete set of visual elements in a chart as $E$.
Specifically, a visual element $e_i$, $e_i \in E$, is defined by a 4-tuple:
\[
e_i = \{ mark, identifier, data, context\} 
  = \{ m_i, id_i, d_i, c_i \},
\]
where $m_i$ denotes the visual mark presented on the chart. The \textit{group-level} and \textit{element-level} marks illustrated in Fig~\ref{fig:granularity} are examples. 
The unique identifier is represented as $id_i$, which enables precise referencing of visual elements in textual representation.
The data values bound to the element are represented by $d_i$, and $c_i$ represents its semantic context, as exemplified in Table~\ref{tab:visual_encoding}.}

\begin{figure}[b]
  \centering
  \includegraphics[width=0.99\linewidth]{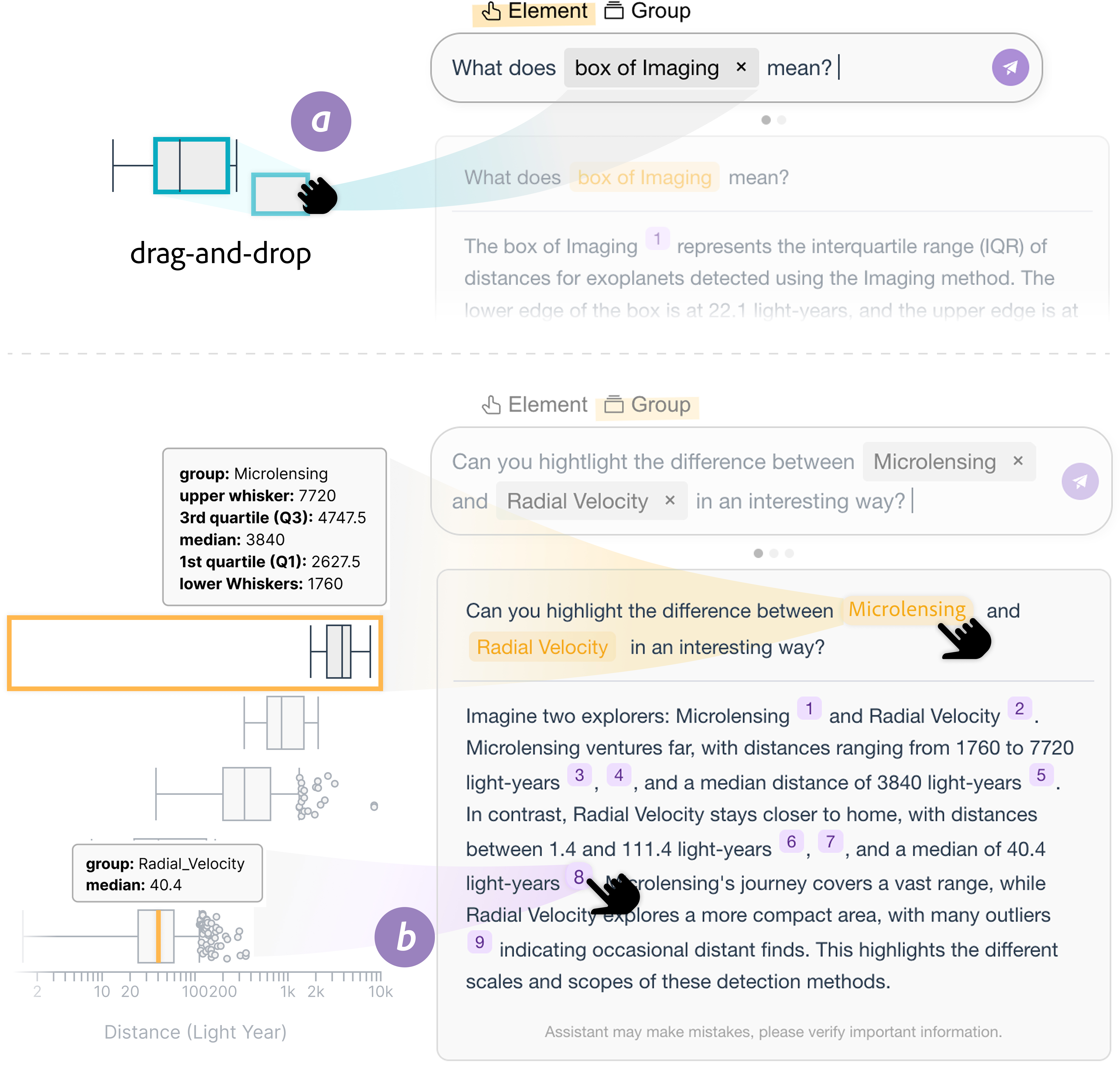}
  \caption{\label{fig:interface}
           The visual-lexical fusion design in \tool includes two aspects: 
           (a) readers can reference element-level and group-level visuals into queries with drag-and-drop, and 
           (b) textual explanations feature inline citations that highlight corresponding visual elements and tooltips.
    }
\end{figure}

\noindent
\textbf{Incorporating visual elements in query.}
\tool allows readers to reference visual elements with drag-and-drop manipulations, eliminating ambiguity when referring to elements during queries.
\revised{Readers can perform the manipulations within the mark space $M = \{m_i | e_i \in E\}$, which encompasses the entire chart area.}
Fig.~\ref{fig:interface}a demonstrates how a reader incorporates a \textit{element-level} mark in query. 
By switching granularity modes, readers can similarly interact with visual elements at \textit{group-level}.
\revised{A dragging event on mark $m_i$ triggers element identification, retrieving visual element $e_i$ with its identifier $id_i$ and bound data $d_i$.
The outcome of the drag-and-drop action is represented as a tag in the messages box, which contains rich content that could identify visual element $e_i$, as shown in the following syntax:}

\noindent\begin{minipage}{\linewidth}
\centering
\vskip -2pt
[tag: [id: $id_i$, data: $d_i$]]
\vskip -2pt
\end{minipage}

\noindent
\textbf{Showing visual details with inline citations.}
Matching textual descriptions to specific elements within a visualization can be time-consuming and difficult in complex visualization for chart readers~\cite{srinivasan2018augmenting,Ottley2019TheCC}.
We design inline citations for \tool to instantly highlight specific visualization elements that match the textual explanation, thereby facilitating \revised{data fact checking}.
Inline citations are shown as labels \includegraphics[width=0.30cm]{src/fig/icon/visualanchor.jpg} (i=1,2,3...), which equally support elements at both \textit{element-level} and \textit{group-level} granularity.
As illustrated in Fig.~\ref{fig:interface}b, readers could hover over these labels to highlight related visual marks in the visualization and display the associated tooltip to verify the accuracy of the textual explanations and quickly recall information from the conversation log.
\revised{To start with, we design a syntax for the inline citation and develop a semantic-aware conversational agent that could invoke the citations when generating explanatory text, as detailed in \S\ref{ssec:citation-tutorial}. 
For a visual element $e_j \in E$ with identifier $id_j$, the agent appends the citation syntax when generating natural language explanations related to its semantic context $c_j$. 
The citation syntax is as follows:} 

\noindent\begin{minipage}{\linewidth}
\centering
\vskip -2pt
[cite: $id_j$]
\vskip -2pt 
\end{minipage}
\revised{
The inline citation syntax is detected and matched with regular expressions from the original responses, and rendered as interactive HTML tags that support hovering interaction.
When readers hover over the label for $e_j$, \tool calls functions to highlight the visual mark $m_j$ and show the corresponding tooltip.}

\subsection{Semantic-Aware Conversational Agent}
\label{ssec:citation-tutorial}
\begin{figure}[t]
    \centering
    \includegraphics[width=0.99\linewidth]{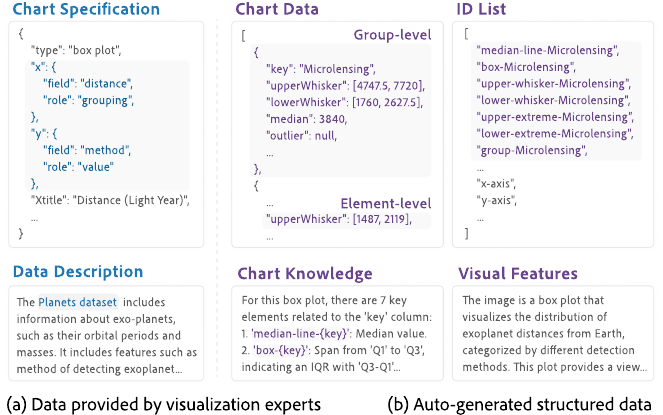} 
    \caption{
    \label{fig:agent}
    The initialization of the semantic-aware conversational agent utilizes multi-source structured data:  (a) visualization specification created by experts, along with the provided contextual data, and (b) auto-generated chart data, ID list, \revised{chart knowledge}, and visual features for information enhancement.
    \vspace{-3mm}
    }
\end{figure}
\label{ssec:agent}

The conversational agent in \tool leverages multi-source structured data from visualization generation and contextual information provided to enhance contextual and fine-grained understanding (Fig.~\ref{fig:agent}).
The backbone of the agent is based on LLMs rather than vision-language models (VLMs), as VLMs consistently struggle with tasks requiring precise spatial information~\cite{rahmanzadehgervi2024vision}.
This also hampers the capacity to address just-noticeable-difference challenges in chart question answering, where high recognition accuracy is essential~\cite{zeng2024advancing}.
To initialize the agent, we align the semantic meaning of visuals (\S\ref{ssec:granularity_semantic}) through prompt engineering, and navigate it through the following chart information.

\begin{figure*}[tbp]
    \centering
    \includegraphics[width=.99\linewidth]{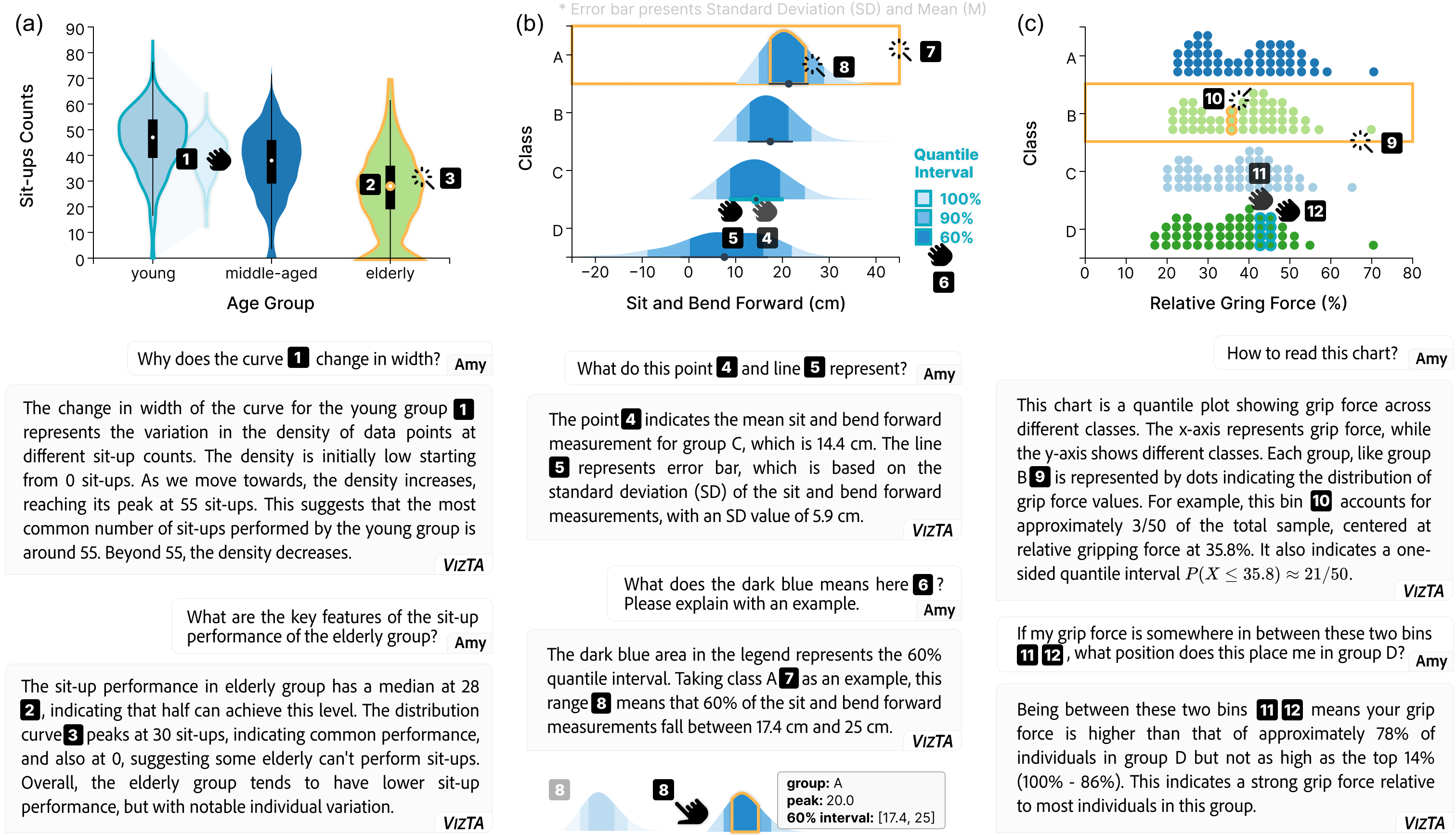} 
    \caption{
    \label{fig:casestudy}
    This figure illustrates a usage scenario in which Amy, using \tool to aid in chart comprehension. 
    Her interaction with these visualizations and the visual feedback from \tool are shown by \includegraphics[width=0.62cm]{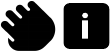} and \includegraphics[width=0.62cm]{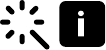}. 
    \vspace{-3mm}
 }
\end{figure*}

\noindent
\textbf{Chart Specification}
refers to the grammar used in initializing a visualization. It includes the chart-type settings and more detailed parameters, serving as an abstraction of its fundamental structure. 
As shown in Fig.~\ref{fig:agent}, the specification identifies the chart type as a boxplot, with two axes representing the grouping by method and the value of distance, respectively.

\noindent
\textbf{Data Description}
provides the metadata of the data shown, serving as a key to help the agent effectively grasp the contextual background of the visualization.
This description can often be obtained from the download source of the data, and visualization experts have the flexibility to include or omit this information based on their needs.
For instance, in Fig.~\ref{fig:agent}a, the data description provides the metadata of the exoplanet dataset.

\noindent
\revised{\textbf{Chart Knowledge} is a collection $C = \{c_i | e_i \in E\}$ that aggregates the semantic contexts of all visual elements presented on a specific type of chart. 
It serves as a structured knowledge base that enables the agent to access and reason about the meanings of visual elements at different granularity levels, facilitating explanations that are visually accurate and semantically meaningful.}

\noindent
\revised{\textbf{Chart Data}
refers to the set $D = \{d_i | e_i \in E\}$ of all data values associated with visual elements, providing the agent access to the detailed chart information.}
For data representing \textit{continuous mark} like density area, key features (\eg, peaks, troughs, extent) are extracted rather than using all data points encoded. 

\noindent 
\textbf{Visual Features}
refer to the long description generated by a VLM, which is prompted with the following instruction: \textit{``Describe in detail the chart type, all visual elements, and insights that can be derived from this visualization.''}
Upon creating the visualization, \tool automatically captures a snapshot of the chart as an image and sends it to the vision model. 
The response is then dynamically incorporated into the system prompt.

\revised{To enable the effective generation of precise visual references, we initialized the LLM-based agent with a citation tutorial prompt.}
This prompt includes carefully designed few-shot examples that demonstrate various citation scenarios, teaching the agent to properly reference and highlight multi-granular visual elements.

\noindent
\revised{\textbf{ID List}
is a collection of all visual element identifiers within the chart, represented as the set $I = \{id_i | e_i \in E\}$.
This list is automatically compiled by retrieving all the \textit{element-level} and \textit{group-level} visual elements in the SVG-formatted chart after initialization.}

\subsection{Implementation}

We implement the system using JavaScript and vue.js for the interface, D3.js for the interactive visualization library, and Tagify.js~\cite{tagify} to transform the visual object into tag input in textarea.
The semantic-aware conversational agent and the VLM for visual features in \tool are based on the ``gpt-4o'' model, accessed through the OpenAI Chat API~\cite{openai2024}.
The responses are streamed word-by-word to achieve nearly real-time replies.

\section{USAGE SCENARIO}
To provide a better sense of how \tool works, we present a usage case involving Amy, an imagined physical educator who is using \tool \revised{to read and interpret visualizations} (Fig.~\ref{fig:casestudy}).
These visualizations \revised{were pre-created by another visualization designer with \tool and published on a web page}, using a dataset of physical fitness measurements provided by the National Sports Administration. 
Amy aims to gain comprehensive data understanding to formulate personalized teaching plans for students of varying ages. 

\revised{
Amy explored the first chart showing sit-ups count among three age groups, as shown in Fig.~\ref{fig:casestudy}a.
While Amy's prior knowledge helps her grasp some elements, she is uncertain about the meaning of the curve  \includegraphics[width=0.30cm]{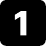}.
Amy queried ``Why does the curve change in width'', and refered to the curve \includegraphics[width=0.30cm]{src/fig/icon/magic1.pdf} with \textit{element-level} drag-and-drop manipulation \includegraphics[width=0.32cm]{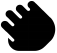}. 
After reading the explanation provided by \tool, Amy understood that the curve represents density and that the sit-ups performance for the young group has a mode of 55. 
Observing the potential age effect, Amy then used \tool to aid in analyzing the sit-up count characteristics of the elderly group.}

\revised{When reading the density plot showing sit and bend forward among different groups (Fig.~\ref{fig:casestudy}b), Amy queried about the explanation of the point \includegraphics[width=0.30cm]{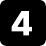} and the line \includegraphics[width=0.30cm]{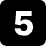}. 
After asking follow-up question about the legend, Amy hovered on label \includegraphics[width=0.30cm]{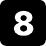} in the response, which highlighted a truncated density area in group A and displayed a tooltip. 
This helped Amy verify and understand the interval area.}

\revised{Amy reviewed the third chart (Fig.~\ref{fig:casestudy}c) comparing relative grip strength across different groups. Confused by the chart at first glance, she accepted the suggested prompt, \textit{``How to read this chart?''}.
After understanding the meaning of a single bin, Amy continued asking questions to quickly assess how her own grip strength would compare using the chart.}

\revised{Throughout this process, Amy explored different types of distributional visualizations while using various \tool interactions. 
As a result, she gained more confidence in explaining these data and developed a clearer understanding of the performance patterns across these three types of physical exercises.}
\section{USER STUDY}

We conducted a between-subjects user study to assess the usefulness and usability of \tool in aiding comprehension of distribution visualizations. 
The evaluation had two main goals: 
(1) assess whether readers could better complete basic comprehension tasks with distributional visualizations using \tool, 
(2) examine whether \tool aids readers in reasoning with distributional visualizations when sharing insights.
We introduce the study setting in \S\ref{ssec:study_setup} and present our findings in \S\ref{ssec:result}.

\subsection{Study Setup}
\label{ssec:study_setup}
\subsubsection{Participants}
The study employed a between-subjects design, as the clear learning effect of the interfaces ruled out a within-subjects approach. 
Potential participants were recruited on campus and screened using Mini-VLAT~\cite{pandey2023mini}, with a passing score of 6 out of 12 to ensure they had a basic understanding of simple visualizations.
From this pool, 24 participants (12 males, 12 females) were selected and divided into two groups: \tool group \smallpar{A1-A12, 6 males and 6 females, $M_{\text{Mini-VLAT}} = 8.5$, SD=2.7}, and \textsc{Baseline} group \smallpar{B1-B12, 6 males and 6 females, $M_{\text{Mini-VLAT}} = 8.5$, SD=1.1}.
The two groups were balanced in terms of gender and Mini-VLAT performance
and assigned to either the grounded or ungrounded condition.
All participants had normal color vision and were either undergraduates or had completed undergraduate education, ensuring a baseline level of numeracy.
Participants were largely unfamiliar with the statistical visualization involved in the reading tasks.

\subsubsection{Conditions}
The two conditions set in the study are as follows:

\begin{itemize}

\item \tool: The system as described in \S\ref{sec:vista}.
\item \textsc{Baseline}: 
\revised{Since our goal is to investigate whether the visual-lexical fusion design can effectively assist chart comprehension, the \textsc{Baseline} design is derived from \tool itself, with the visual-lexical fusion interaction design being ablated. 
Correspondingly, during the agent initialization phase, the associated information sources, including chart data ($D$), chart knowledge ($C$), and ID List ($I$) are also removed.}

\end{itemize}

To ensure the two conditions are consistent, three considerations are made: 
(1) The basic interface design for \textsc{Baseline} is consistent with \tool. 
(2) Both systems use identical prompts besides the removed enhanced information.
(3) They used the same model and temperature setting (i.e., gpt-4o, 0.2), accessed through the OpenAI Chat API and support for stream generation. 

\subsubsection{Scenario Design and Tasks.}
We crafted six reading scenarios, each accompanied by designed questions to form the reading tasks.
This collaborative effort involved two experienced visualization educators.
Detailed scenarios and tasks are available in the supplementary material. 

\noindent
\textbf{Scenario Design.}
Each scenario is designed as an independent scene, featuring:
(1) a \textit{title} for the topic,
(2) \textit{background information} for context,
(3) an \textit{interactive visualization} depicting data distribution with data values accessible via tooltips,
(4) a \textit{caption} encapsulating the essence of the visualization and offering supplementary information.
The selected visualization types were box plots, density plots, violin plots, and quantile dot plots, with their presentation formats shown in Fig.\ref{fig:scope}.
The visualizations were created using real-world data, covering a broad range of topics. They all depict statistical variability and uncertainty in the data.

\noindent
\textbf{Task Design.}
We structured our tasks using two formats: 
(1) single-choice questions (SCQs) for the precise assessment of factual knowledge with minimal ambiguity, and 
(2) open-ended questions (OEQs) to delve into reasoning and  identification of misconceptions.
For task one, we selected four scenarios to create SCQs focused on basic comprehension (\textsc{4 scenarios} $\times$ \textsc{4 questions}). 
These SCQs targeted two key areas:
(1) general analytics tasks, such as data retrieval and filtering, and 
(2) chart-specific questions assessing understanding of uncertainty and variability (e.g., \textit{``Which group has the smallest interquartile range (IQR)?''}).
For task two, we designated two scenarios to develop OEQs aimed at oral reporting (\textsc{2 scenarios} $\times$ \textsc{2 questions}). 
These OEQs centered on higher-level tasks involving comparison and summarization.
The OEQs were assessed using a pass rate criterion, where a response was considered a \textit{``pass''} if it addressed the question without demonstrating any misunderstanding of the visual information shown on the chart. 
All questions were pilot-tested to ensure clarity and comprehensibility before the experiment.

\subsubsection{Procedure}
During the study, participants were assigned to either \tool and \textsc{Baseline} conditions and tasked with completing the assigned tasks to the best of their ability.
They were unaware of their condition group and the features of the tool used in the other condition.
To streamline the experimental process, we developed a website for the user study.
This platform facilitated the presentation of visual materials and systematically recorded the interaction event from participants (e.g., answers to questions, prompting, clicks, and event time).
The screen and microphone were recorded.
The user study was conducted in a one-on-one, in-person format, and lasted approximately one hour.
In appreciation for their time, participants received the equivalent of \$7.

\noindent
\textbf{Introduction and Tutorial}.
After signing the consent form and completing the demographic questionnaire, participants were introduced to the background of this study. 
They then reviewed tutorial documentation specific to their assigned condition, with a hands-on trial using an example chart to familiarize themselves with the tool. 

\noindent
\textbf{Task One: Basic Comprehension} (\textsc{4 scenario}).
Participants sequentially engaged with four scenarios, each allowing a two-stage process: 
(1) \textit{Exploration}: they had 150 seconds to interact and understand all visual elements of each chart. 
(2) \textit{Question Answering}: answering four single-choice questions in 40 seconds each, with guidance to skip rather than guess when feeling uncertain. 

\noindent
\textbf{Task Two: Reasoning in Oral Sharing} (\textsc{2 scenario}).
For task two, participants were shown two scenarios, each with one chart accompanying two questions.
For each question, participants went through two stages:
\textit{(1) Exploration}: They had two minutes to explore the visualization with the reading assistant and prepare their responses.
\textit{(2) Oral Sharing}: Participants then had one minute to present their answers by speaking out loud.
They were encouraged to deliver high-quality responses by focusing on the contextual meaning of the visualizations, highlighting insights with specific data values, and sharing unique findings relevant to the questions.

\noindent
\textbf{Questionnaire Survey}.
After completing all the tasks, participants were asked to complete a questionnaire that included NASA-TLX and System Usability Scale (SUS). 

\noindent
\textbf{Semi-structure Interview}.
At the end of the user study, participants are invited to discuss their visualization reading experience and provide feedback on the reading assistant. 
They were encouraged to offer design suggestions, voice concerns, and share memorable experiences.

\begin{figure*}[tbp]
    \centering
\includegraphics[width=0.95\textwidth]{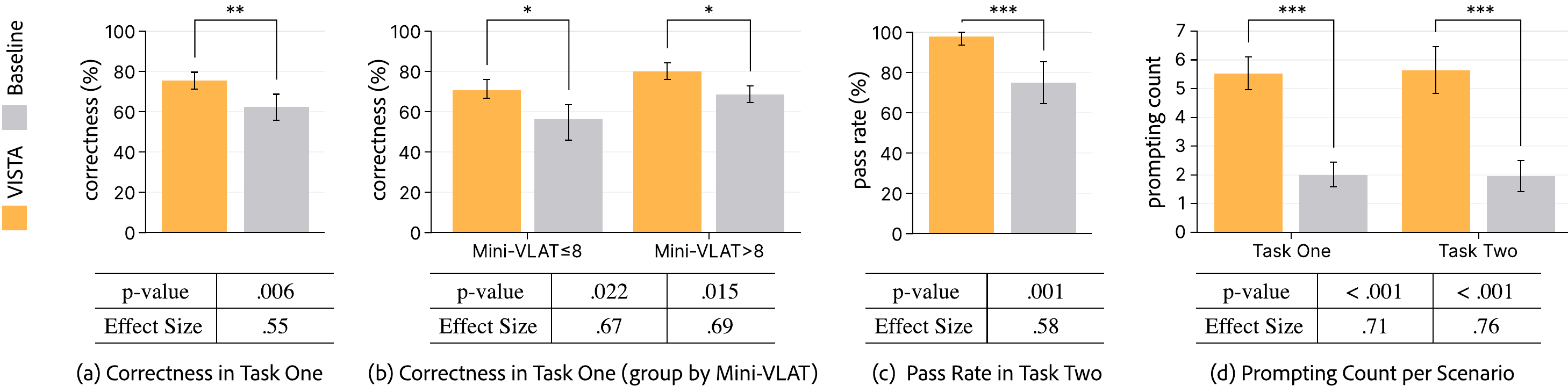}
    \caption{\label{fig:result}Results to (a) correctness in task one, (b) correctness in task one grouped by Mini-VLAT performance, (c) pass rate in task two, and (d) prompting count per scenario grouped by task. Error bars represent 95\% CIs, and significance values are reported for p <
.05 (*), p < .01 (**), and p < .001(***).
    }
    \vspace{-5mm}

\end{figure*}

\subsection{Result}
\label{ssec:result}

We present our result by first sharing the findings of the two evaluation goals: 
 (1) \tool users showed higher accuracy in basic comprehension tasks and benefited across varying visualization literacy levels. 
(2) \tool users made fewer errors, had a higher passing rate in reasoning tasks, and more often cited precise data points from visualizations to bolster their conclusions.
These findings are corroborated by the quantitative results in \S\ref{sssec:quantitative_result}, and further feedback from post-interviews is detailed in \S\ref{sssec:qualitative_result}.

\subsubsection{Quantitative Results}
\label{sssec:quantitative_result}
We ran the Mann-Whitney U test for non-parametric analysis and calculated 95\% CIs with the bootstrap method.
Two authors independently coded the answers for open-ended questions. Any discrepancies in coding were resolved through discussion between the authors.
Overall, participants had better results with \tool on objective metrics (correctness, pass rate, interaction used) and subjective metrics (SUS).

\noindent
\textbf{Correctness of Single-choice Question.}
Mann-Whitney U test \smallpar{$n_1 = n_2 = 12$} found a significant effect of the condition on correctness \smallpar{$p=.006$, $r=.55$}, as shown in Fig.\ref{fig:result}a.
Participants using \tool\ had higher correctness rates \smallpar{$M = 75.5$, 95\% CI [71.4, 79.7]} than those in the \textsc{Baseline} group \smallpar{$M = 62.5$, 95\% CI [55.7, 68.2]}.
The effect of condition was consistent across participants with both lower \smallpar{Mini-VLAT \( \leq 8 \), $n_1 = n_2 = 6$, $p=.022$, $r=.67$} and higher \smallpar{Mini-VLAT \( > 8 \), $n_1 = n_2 = 6$, $p=.015$, $r=.69$} visualization literacy, Fig.\ref{fig:result}b.
Participants with lower visualization literacy in \tool\ \smallpar{$M = 70.8$, 95\% CI [$66.7$, $76.0$]} almost caught up those higher in \textsc{Baseline} \smallpar{$M = 68.8$, 95\% CI [$64.6$, $72.9$]}.

\noindent
\textbf{Reporting Quality.}
Participants in \tool\ \smallpar{$M = 97.9$, 95\% CI [$93.8$, $100.0$]} significantly outperformed \textsc{Baseline} \smallpar{$M = 75.0$, 95\% CI [$64.6$, $85.4$]} in reporting and reasoning tasks\smallpar{$p=.001$, $r=.58$}. 
In fact, only one participant in \tool failed to report the correct answer to one question. 
In task two, \tool\ participants cited 150 meaningful data points to support their conclusions, compared to 84 by \textsc{Baseline} participants (\(p = .006\), \(r = .18\)). 
Specifically, \tool\ participants reported 117 precise values and 33 estimated values, whereas \textsc{Baseline} participants reported 31 precise values and 53 estimated values.

\noindent
\textbf{Interaction Depth \& Rating.}
With \tool, participants tended to prompt more to the system for conversation than \textsc{Baseline}, and significant effects could be observed both in scenario of task one \smallpar{$p<.001$, $r=.71$, $M=5.5$ vs. $M=2.0$} and task two \smallpar{$p<.001$, $r=.76$, $M=5.6$ vs. $M=2.0$}. 
The two aspects of the visual-lexical fusion interaction designs of \tool received mostly positive ratings across five dimensions: novelty, helpfulness, ease of use, fun, and focus.
Among these, \textit{inline citation} is recognized for its ability to help users maintain focus during tasks \smallpar{$M=6.92$, $SD=.28$}.

\noindent
\textbf{System Usability \& Workload.}
\revised{On the System Usability Scale (SUS),  participants rated \tool higher overall \smallpar{M=5.91 vs. M=5.4 on a 7-point scale}.}
It showed a significant effect in three dimensions, and \tool was rated better in \textit{``easy to use (Q2)''} \smallpar{$p=.044$, $r=.40$}, \textit{``various functions were well integrated (Q3)''} \smallpar{$p=.049$, $r=.71$}, and \textit{``could imagine people would learn to use this system very quickly (Q4)''} \smallpar{$p=.046$, $r=.37$}. 
For workload report in NASA-TLX, a significant effect is found on performance \smallpar{$p=.030$, $r=.43$}.
Compared to \textsc{Baseline}, participants using \tool reported higher perceived performance \smallpar{$M=5.4$ vs. $M=4.4$}.

\subsubsection{Qualitative Results}
\label{sssec:qualitative_result}

\textbf{\tool offers a user-friendly interaction design, enabling readers to seek guidance efficiently.}
Almost all participants in the \tool group praised the tool in aiding their comprehension of distributional visualizations. 
They enjoy the following functionalities designed.
\uline{Incorporating visual elements in query} simplifies the process of posting questions by
\textit{``easier to describe specific intent''} (A8), and offers a description method that is accurate and could \textit{``dive into details''} (A11).
Participants also enjoyed it from an interactive aspect, with A10 commenting on the process as
\textit{``cute and fun''}.
However, two users expressed concerns about this interaction for \textit{``the complexity of drag-and-drop operations''} (A2) and noted that sometimes \textit{``forgot this interaction due to unfamiliarity''} (A1).
Participants agreed that \uline{inline citation} facilitates seamless integration of text and graphics, as A8 commenting \textit{``the visual guidance make it easy for me to know where to focus [...] and based on the number of labels, I can quickly grasp how much information is being referenced''}.
Interestingly, users mentioned that contextual explanation of the chart provided by \tool made the visualization content memorable, 
\textit{``Almost like telling a story. I even hadn't grasped that understanding on my own''} (A5), and \textit{``lower the barrier to read these scientific and complex visualizations''} (A11).

\textbf{In conversational interfaces designed for visualization, users have a stronger willingness to discuss chart contents when the explanation and visual aids are more contextual and detailed.}
When visualizations appear in a conversational interface, users naturally expect the system to understand the details of the visualization and want to engage in discussions of it.
In line with this, all users inquired about the content of the charts in their initial queries. 
During the interview, we asked participants why they interacted more or less with the assistant.
Users from the baseline group highlighted two significant challenges: the need to meticulously frame their questions and the effort required to interpret the responses from the chatbot within the context of the visualization. 
Reasons for reduced interaction also included disappointment, as comment from B7, \textit{``Addressing these questions requires precise values, but it can’t provide that level of detail. It offers guidance on reading the chart, but lacks specific examples''}. 
Conversely, \tool participants experienced a more contextually grounded discussion, noting that A6 \textit{``I don't need to struggle at thinking about how to ask''}.
This facilitated an environment where users felt understood from the initial explanation, as participant A7 stated, \textit{``It pointed out elements in chart and explained, I felt it truly understood this chart. This made me want to engage more deeply in the conversation''}

\textbf{Communication serves as a remedy for misunderstanding, it facilitates quicker error identification and provides means for verifying understanding in chart.}
\label{sssec:communication}
An interesting finding is the role of conversational interaction in reshaping and enhancing comprehension of visualizations.
\tool users, through these interactions, not only pinpointed misconceptions, such as \textit{``I overlooked a detail that a dot (in quantile dotplot) is not a sample; it represents probability''} (A5), but also deepened their grasp of key concepts, exemplified by (A3) \textit{``it reminded me that the median is resistant to the influence of outliers, broadening my understanding.''}
When inquiring why baseline group users had low confidence in their understanding, one typical response was \textit{``I don't know how to verify if my understanding is correct''} (B10). 
In contrast, a \tool user (A12) expressed, \textit{``I can easily confirm my understanding, like having a companion that lightens my load''}.
However, we also observed that some \tool users with lower visualization literacy tended to rely heavily on the system guidance to complete tasks, highlighting the need to encourage more independent thinking.
\section{DISCUSSION, LIMITATION, and FUTURE WORK}

\textbf{Possibility for visualization education to a broader audience.}
The participants we recruited were already well-versed in data literacy, as they were mostly graduate students from a research university, with a few undergraduates. 
\revised{We did not test \tool with individuals with lower data or visualization literacy. 
Further efforts are required to validate whether and how such interaction paradigms could benefit them and what adaptations are necessary.}
We envision that the interactive reading modules created by \tool could make visualization education more accessible to a broader audience of general readers.
\revised{
While our study assessed chart comprehension efficiency with AI collaboration, we did not adequately measure their reliance levels on AI support or the long-term learning effects.
Another question highlighted by our work is the adaptive balance of guidance: excessive guidance may inhibit creative thinking~\cite{shu2024particular} or lead to over-reliance, while insufficient guidance could confuse novice chart readers. 
}

\noindent
\textbf{LLMs do make mistakes in mathematical reasoning.}
Limitations in LLMs' mathematical reasoning abilities and their tendency to hallucinate indeed impact the performance of \tool.
In our testing and user experiments, \tool rarely made mistakes when reporting precise data values or using inline citations. 
However, errors could occur when tasks require further reasoning based on those values. 
To optimize the use of LLMs in visualization comprehension, several considerations are necessary. 
First, guide users to develop appropriate mental models of AI capabilities, preventing undertrust or overtrust~\cite{yang2020how}.
Second, improving mathematical reasoning abilities through strategies like CoT, tool augmentation~\cite{chen2023good}, or multi-agent planning~\cite{Li2024AgentOrient}, could enhance system reliability.
\revised{Third, for scenarios with multiple charts or complex data contexts, retrieval-augmented generation (RAG) may help maintain accuracy by retrieving precise values rather than relying solely on LLMs. 
Additionally, future work should include quantitative evaluation of mathematical fact descriptions and visual citations across different LLMs.}

\noindent
\textbf{From pre-designed interactive lessons to everyday chart creation and reading.} 
Currently, creating educational cases for distributional visualizations with \tool typically requires the involvement of a visualization creator.
Looking ahead, we envision two directions for expansion.
First, to expand the system to support a wider range of chart types beyond distributional visualizations, a more extensive gallery could be developed for educational purposes.
Second, the visual-lexical fusion interaction design and the semantic-aware conversational agent of \tool could be integrated into professional visualization generation tools 
and beyond education, such as storytelling or facilitating professional data analysis.

\noindent
\revised{\textbf{Facilitating visual communication with LLMs.}
As language-driven models, LLMs face inherent challenges in supporting visual communication.
This gap manifests in both understanding the visual references from users and providing appropriate visual feedback.
To accurately capture the communicative intent of users in visualization context, our work implements drag-and-drop interaction to bring visual elements in queries, yet visualization interfaces encompass a broader spectrum of fundamental interactions~\cite{yi2007interaction}. 
Second, enabling LLMs to provide diverse visual feedback deserves further exploration. 
Similarly, our visual feedback currently highlights existing elements, while future efforts could be extended to graphical overlays~\cite{hao2024finflier} or augmentations~\cite{shi2025augementing} for analytical tasks in conversation interaction.
}
\section{CONCLUSION}

Interpreting distributional visualizations presents a significant challenge for general readers.
While LLMs are increasingly recognized as effective tools for visualization education, effective approaches for visualization-centered conversations in conversational interfaces remain underexplored.
Inspired by a formative study on the demands for a chart teaching assistant, we developed \tool, a visualization teaching assistant designed to aid readers in comprehending reasoning with distributional visualizations.
\tool features the design of \textit{visual-lexical fusion interaction} to meet reader intent, and a \textit{semantic-aware conversational agent} capable of explaining details in visualization. 
A between-subjects study assesses whether readers with \tool could better complete basic comprehension tasks and reason
with distributional visualizations.
The evaluation results suggest that the designs in \tool help readers perform better and receive more favorable feedback.

\noindent
{\textbf{Acknowledgment}.} \ 
The authors wish to thank anonymous reviewers for their constructive comments. 
The work was supported in part by National Natural Science Foundation of China (72371217, 62172398, 62472357), the Guangzhou Industrial Informatic and Intelligence Key Laboratory No. 2024A03J0628, the Nansha Key Area Science and Technology Project No. 2023ZD003, and Project No. 2021JC02X191.

\bibliographystyle{eg-alpha-doi} 
\bibliography{reference}
\end{document}